\documentclass[final,3p,times,twocolumn]{elsarticle}
 \biboptions{comma,sort&compress}
\usepackage{here}
 \usepackage{graphicx}
  \usepackage{epsfig}

\usepackage{amsmath}
\usepackage{amssymb}

\newcommand{\nn}{\nonumber}
\newcommand{\be}{\begin{equation}}
\newcommand{\ee}{\end{equation}}
\newcommand{\bea}{\begin{eqnarray}}
\newcommand{\eea}{\end{eqnarray}}
\def\al{\alpha}

\def\siml{{\ \lower-1.2pt\vbox{\hbox{\rlap{$<$}\lower6pt\vbox{\hbox{$\sim$}}}}\ }}

\newcommand{\ord}{{\cal O}}

\def\msb{{\overline{\rm MS}}}

\def\nin{\noindent}
\def\beq{\begin{equation}}
\def\eeq{\end{equation}}
\def\bea{\begin{eqnarray}}
\def\eea{\end{eqnarray}}

\journal{Nuc. Phys. (Proc. Suppl.)}

\begin{document}

\begin{frontmatter}



\title{The QCD static potential in 2+1 dimensions at weak coupling}

 \author[label1]{Maximilian Stahlhofen\corref{cor1}}
  \address[label1]{Grup de F\'\i sica Te\`orica and IFAE, Universitat
Aut\`onoma de Barcelona, E-08193 Bellaterra, Barcelona, Spain.}
\cortext[cor1]{Speaker}
\ead{stahlhofen@ifae.es}



\begin{abstract}
\noindent
Using the effective theory pNRQCD we determine the potential energy of a color singlet quark-antiquark pair 
with (fixed) distance $r$ in three space-time dimensions at weak coupling 
($\al\, r \ll 1$). The precision of our result reaches ${\cal O}(\al^3r^2)$, i.e. NNLO in 
the multipole expansion, and NNLL in a $\al/\Delta V$ expansion, where 
$\Delta V \sim \al \ln(\al r)$. We even include all logarithmic terms up to $\rm N^4$LL order 
and compare the outcome to existing lattice data.
\end{abstract}

\begin{keyword}


static potential \sep 2+1 dimensions \sep pNRQCD \sep ultrasoft corrections

\end{keyword}

\end{frontmatter}


\section{Introduction}
\nin
The potential energy of a static quark-antiquark pair in the color singlet at short distances $r$ is an essential ingredient in the theoretical description of heavy quarkonium.
Its large-distance behavior, which is probed e.g. in lattice simulations, indicates confinement.
The determination of the static potential in three space-time dimensions (3D) represents an important check of the methods used for the four-dimensional (4D) calculation~\cite{FSP,Anzai:2009tm,Smirnov:2009fh}. The results can also be applied within thermal QCD.
In this paper we determine the 3D static potential for $\alpha r \ll 1$ using the effective field theory ``potential nonrelativstic QCD'' (pNRQCD)~\cite{Pineda:1997bj,Brambilla:2004jw}.
Unlike conventional perturbative QCD, this effective theory framework allows for so-called ultrasoft effects, which are crucial for consistent results beyond one loop~\cite{Appelquist:1977es,short,Brambilla:1999xf}.
We discuss the renormalization group structure of the 3D static potential and present recent results up to ${\cal O}(\alpha^3 r^2)$ and partly N$^4$LL precision. We also compare these results to existing lattice data. The work presented here is based on Ref.~\cite{OurStatPot3D}.


\section{Theoretical preliminaries}
\nin
Because in $D$ dimensions the mass dimension of the coupling is $[g^2]=M^{4-D}$, $g^2r^{4-D}$ is a dimensionless (expansion) parameter and we have (at least) the following physical scales involved in the problem: $1/r$ (soft), $V \sim 1/r \times g^2r^{4-D}$ (ultrasoft), $g^{\frac{2}{4-D}}$ (non-perturbative).
In order for perturbation theory at the soft scale and the pNRQCD multipole expansion to make sense we demand 
$g^2r^{4-D} \ll 1$, i.e. weak coupling.\\
For $D=3$ at short distances we find $V \sim g^2\ln(r\nu)$, where $\nu$ is the pNRQCD renormalization scale. This implies that the ultrasoft expansion parameter $g^2/V \sim 1/\ln(r\nu) \ll 1$, if we choose $\nu \sim  V$. We conclude that we can use perturbation theory at the ultrasoft scale $V$. Therefore we formally distinguish between
the scales $V$ and $g^2$. Logarithms from the ultrasoft perturbative computation will have the form $\ln ( V/\nu)$ and are rendered small, if we set $\nu \sim  V$. 
Thus it is legitimate to consider the ultrasoft regime as perturbative, i.e. the pNRQCD loop expansion makes sense (for sufficiently small $r$).\\
In the following we will use the index ``$B$'' to explicitly denote bare quantities. Parameters without this index are understood to be renormalized in the $\rm MS$ scheme. 
Throughout this paper we will use the notation $D=3+2 \epsilon$.
In position space the bare singlet potential can be schematically written as
\begin{align}
\label{VsBD}
V_{s,B}
\equiv&
-C_Fg_B^2\sum_{n=0}^{\infty}\frac{g_B^{2n}c_n(D)r^{-2(n+1)(\epsilon-\frac12)}}{r}\,.
\end{align}
The singlet static energy can be considered to be an observable for our purposes. 
It consists of the potential, which is a pNRQCD Wilson coefficient, and an ultrasoft 
contribution\footnote{If one has enough precision also non-perturbative effects at the scale $g^2$ should be included. We will address 
them in Sec.~\ref{results}.}, either bare or renormalized:
\be
E_s(r)=
V_{s,B}+\delta E^{us}_{s,B}
=
V_{s}+\delta E^{us}_{s}
\,.
\label{Estot}
\ee
The soft contribution $V_{s,B}$ equals the purely perturbative bare static potential, which was computed in Ref.~\cite{Schroder:1999sg} up to two loops, i.e. $\ord(\al^3 r^2)$. 
It is IR divergent at this order.
Using pNRQCD in the static limit the ultrasoft contribution can be expressed in a compact form at NLO in the multipole expansion (but exact to any order in the ultrasoft loop expansion) through the 
chromoelectric correlator. It reads (in the Euclidean)
\begin{align}
\delta E^{us}_{s,B}
  =&\, V_A^2 \frac{T_F}{(D-1) N_c}\, {\bf r}^2 \int_0^\infty \!\! dt  e^{-t\,\Delta V_B} \nn\\[1 ex]
&\times \langle vac| g_B {\bf E}_E^a(t) \phi_{\rm adj}^{ab}(t,0) g_B {\bf E}_E^b(0) |vac \rangle
\,,
\label{deltaVUS}
\end{align}
where we have defined $\Delta V \equiv V_o-V_s$.
The concrete result for the ultrasoft correction in $D$ dimensions up to two loops is given in Ref.~\cite{OurStatPot3D}.
It is known at one loop since Ref.~\cite{Pineda:1997ie} (see also~\cite{short,KP1}) and was deduced at two loops from 
the results obtained in Refs.~\cite{Eidemuller:1997bb,Brambilla:2006wp}.\\
%
The renormalized coupling constant $\alpha = \frac{g_B^2\nu^{2\epsilon}}{4\pi}$ has integer mass dimension 
and does not run in three dimensions as a consequence of the super-renormalizability of the theory.\\
The bare potentials $V_B$ in position space also have integer mass dimensions and following Ref~\cite{RG} we define
\be
\label{VBsplitting}
V_B=V+\delta V\,.
\ee
$\delta V$ will generally depend on $\al$ and $V$. In the MS renormalization scheme it takes the form
\be
\delta V
=
\sum_{n=1}^{\infty} \, \frac{Z^{(n)}_{V}}{\epsilon^n}\,,
\ee
from which 
we can derive the RGE's for the different renormalized potentials $V$ in the usual way.
%
%
%
%
\begin{figure}
\begin{center}
\includegraphics[width=0.2 \textwidth]{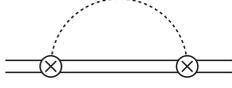}
\end{center}
\caption{\it One-loop contribution to the octet propagator. The dotted line 
represents the $A^0$ component of the ultrasoft gluon field.}
\label{usoctet3D}
\end{figure}\\
In pNRQCD at leading order in the multipole expansion the singlet field of the quark-antiquark system is free, i.e. 
it does not interact with gluons. It is therefore renormalization scale independent at $\ord(r^0)$.
Similarly the singlet potential is not renormalized at this order:
\be
\delta V_s={\cal O}(r^2)
\,.
\label{LOdeltVs}
\ee
For the octet field the situation is different. Even at leading order in the multipole expansion it has 
a residual interaction with ultrasoft gluons. 
The octet potential receives an ultraviolet (UV) divergent correction from 
the one-loop self-energy diagram shown in Fig. \ref{usoctet3D}:
\be
Z^{(1)}_{V_o}
=
\frac{C_A}{2}\, \al + {\cal O}(r^2)
\,.
\label{ZVo}
\ee
Higher loop diagrams cannot contribute at $\ord(r^0)$.
This is because the potentials must appear perturbatively (with positive powers) 
in the Z's. Since $\al$ has positive mass dimension, the potentials would appear with negative powers 
in higher loop corrections to Eq.~(\ref{ZVo}), which 
is not allowed by renormalizability. By the very same reason the octet field does not require renormalization at $\ord(r^0)$. 
With a similar argument and Ref.~\cite{Brambilla:2006wp} we find $V_{A/B}=1+{\cal O}(\al^2)$.\\
%
Solving the corresponding RGE's and matching to the soft tree-level result we find from Eqs.~\eqref{LOdeltVs} and~\eqref{ZVo}:
\begin{align}
\label{VoRG}
\Delta V(\nu)
=-\al C_A\ln(r\,\nu\, e^{\gamma_E/2}\sqrt{\pi}\,)+\ord(r^2)+\ord(\epsilon)\,.
\end{align} 
We now focus on the renormalization of $V_s$ beyond ${\cal O}(r^0)$. 
The singlet potential is IR safe up to soft one-loop order. At two soft loops in dimensional regularization IR poles up to $\ord(1/\epsilon^3)$ appear~\cite{Schroder:1999sg}.
The ultrasoft computation in pNRQCD, i.e. Eq.~\eqref{deltaVUS}, yields the counter\-terms 
\begin{align}
Z^{(1)}_{V_s}
&= r^2 \Delta V^2 \al\, Z^{(1,1)}_{V_s} + r^2 \Delta V\al^2 \,Z^{(1,2)}_{V_s} + r^2\al^3\, Z^{(1,3)}_{V_s} , \label{Z1Vs}\\
Z^{(2)}_{V_s}
&= r^2 \Delta V \al^2\, Z^{(2,1)}_{V_s} + r^2\al^3\, Z^{(2,2)}_{V_s} , \label{Z2Vs}\\
Z^{(3)}_{V_s}
&= r^2\al^3\, Z^{(3,1)}_{V_s} , \label{Z3Vs}\\
Z^{(n)}_{V_s}
&=0 \quad \forall \quad n>3\, , \label{ZnVs}
\end{align}
where the explicit expressions for the $Z^{(x,y)}_{V_s}$ are given in Ref.~\cite{OurStatPot3D}.
These are the complete  ${\cal O}(r^2)$ results. Eq.~\eqref{ZnVs} reflects the super-renormalizability of the theory. 
Eqs.~(\ref{Z1Vs}-\ref{Z3Vs}) are obtained as follows:\\ 
$Z^{(1,1)}_{V_s}$ comes from the one-loop $1/\epsilon$ divergence in Eq.~(\ref{deltaVUS}) and fixes also $Z^{(2,1)}_{V_s}$ and $Z^{(3,1)}_{V_s}$ by RG consistency.
$Z^{(1,2)}_{V_s}$ follows from the remaining $1/\epsilon$ divergence in the ultrasoft two-loop computation for $\delta V_s$, once all subdivergences have been subtracted. From this result we derive $Z^{(2,2)}_{V_s}$ again by RG arguments.
Because the respective divergent parts of the bare quantities in Eq.~(\ref{Estot}) have to cancel, we can now also determine the missing term $Z^{(1,3)}_{V_s}$ without actually performing the corresponding ultrasoft three-loop calculation~\cite{OurStatPot3D}. This is possible since $Z_{V_s}$ must not contain terms with negative powers of $\Delta V \sim \al\ln (r \nu)$, which cannot be absorbed by the potential, cf. Eq.~(\ref{VsBD}).
The fact that the resulting $Z^{(1,3)}_{V_s}$ is indeed independent of $\Delta V$ is a non-trivial crosscheck of both, the soft and the ultrasoft calculations.
Thus we have found the complete RG structure of $V_s$ at ${\cal O}(r^2)$.

\section{Results}
\label{results}
\nin
From the counterterms determined in the previous subsection, we can derive
the complete running of the singlet static potential at ${\cal O}(r^2)$.
By solving the RG equations we obtain
\be
\label{VsRG}
V_s(\nu)=V_s(r;\nu \!=\! \frac1{r})+V_s^{\rm RG}(r;\nu)\,,
\ee
where $V_s^{\rm RG}(r;\nu)$
is the running and $V_s(r;\nu \!=\! \frac1{r})$
is the initial matching condition, which we have determined using the bare soft data of Ref.~\cite{Schroder:1999sg}. The full MS results are given in Ref.~\cite{OurStatPot3D}.
Note that Eq.~\eqref{VsRG} is the complete RG improved static potential (i.e. the soft contribution to the static energy) up to $\ord(r^2)$.
Adding the finite parts of $\delta E^{us}$ at one and two loops after minimal subtraction to Eq.~\eqref{VsRG} and setting $\nu = \Delta V$ to resum potentially large ultrasoft logarithms we obtain
\begin{align}
E_s(r)
&=
C_F \alpha \ln (r^2 \nu^2_s \pi e^{\gamma_E} 
)   +  \frac{\pi}{4} C_F (7 C_A-4 n_f T_F) \alpha^2 r  \nn
\\
&+C_F \alpha ^3\, r^2
\Bigg\{
\frac{1}{6} C_A^2 \ln ^3(r \Delta V) \nn\\
&+\frac{1}{4} C_A^2 (2 \gamma_E \!-\! 1 \!-\! 2\ln 2 ) \ln ^2(r \Delta V) \nn\\
&+\bigg[ n_f\! T_{\!F} \Big(C_A \big(2 \!+\! \frac{19 \pi ^2}{48}\big) +C_F \big(5 \!-\! \frac{\pi ^2}{2}\big)\Big) - (n_f\!T_{\!F})^2 \frac{ \pi^2}{8} \nn\\
&+C_A^2 \Big(\frac{13 \pi ^2}{384}+ \frac{1}{2} \gamma_E ^2 + \frac{1}{2} \ln ^2 2 -\frac{1}{4}\gamma_E  (1\!+\! 4 \ln 2) \nn\\
&-\frac{1}{4} (11\!+\!\ln \pi ) \Big) \bigg] \ln (r \Delta V) \Bigg\} +{\cal O}(\al^3 r^2 \ln^0)
\,.
\label{NNLL}
\end{align}
This is the full result for the static energy up to ${\cal O}(\al^2r^2)$ and ultrasoft NNLL order
expressed as a double expansion in $\al r$ (multipole) and $1/\ln(r\Delta V)$ (ultrasoft).\\
The omitted $\ord(\al^3 r^2)$ terms do not contain logarithms of $r \Delta V$.
Eq.~(\ref{NNLL}) is invariant under a change of $\nu$ up to $\ord(\frac{\al^4 r^2}{\Delta V})$. 
%
The dependence of $E_s(r)$ at tree level on the factorization scale $\nu_s$ is related to the 3D relic of the 4D pole mass renormalon and would cancel, if we add twice the heavy quark mass to Eq.~\eqref{NNLL}. In this work it will however be of no importance.
Finally we would like to note that the condition 
$\nu \equiv \Delta V(\nu)$, produces a $\nu$ independent scale
that is nonperturbative in $\al$ and resums a certain class of logarithms, see Ref.~\cite{OurStatPot3D}.\\
Since for $\nu \equiv \Delta V(\nu)$ the ultrasoft logarithms vanish, we only have to add the (RG-scheme dependent) constant term
\be
\label{NNNLLmatching}
V_s(r;\nu \!=\! \frac1{r})\Big|_{{\cal O}(\al^3)} + C_F C_A^2r^2 \al^3 c_{2,0}
\ee
to Eq.~\eqref{NNLL} to reach N${}^3$LL order.
$c_{2,0}$ can be computed perturbatively, but requires a three-loop pNRQCD computation which has not been performed yet.\\
At even higher orders in the $\al/\Delta V$ expansion, non-perturbative effects start to contribute. 
In order to study these effects related to loop momenta $k \sim \al$, we integrate out the $\Delta V$ scale. 
This means integrating out the octet field and ultrasoft gluons. The degrees of freedom left are the 
singlet field and nonperturbative gluons with energy and momentum of order $\al$. 
The resulting Lagrangian, including the leading order nonperturbative effects at ${\cal O}(r^2)$, reads
\begin{align}
 {\cal L}_{\rm np} =&
{\rm Tr} \Biggl\{ {\rm S}^\dagger \left( i\partial_0  - V_s(r) -\delta E_s^{us}  \right) {\rm S} \Biggr\} \nn\\
&-  \frac{C_{np}}{\Delta V} {\rm Tr} \left\{  {\rm S}^\dagger {(g\bf E \cdot r})^2 \,{\rm S}\right\}
\label{pnrqcdnp}
\end{align}
for the case without light fermions ($n_f=0$) to which we restrict ourselves in the following.\footnote{
If we were to include light fermions there would also be operators $\propto {\rm S}^\dagger \bar q q \,{\rm S}$.
They could generate corrections to the static energy, due to the quark condensate, which are of the same parametric order as the 
purely gluonic ones.}
The coefficient of the nonperturbative operator in Eq.~(\ref{pnrqcdnp}) is $C_{np}=1$ at leading order in the 
$\frac{\al}{\Delta V}$ expansion. This result is obtained  
by matching to a pNRQCD tree-level diagram, where two gluons couple to the singlet field at $\ord(r^2)$.\\
The interaction with nonperturbative gluons produces a shift of the energy which is proportional to the 3D gluon 
condensate:
\begin{align}
\delta E^{np}_{s,B}&=\frac{r^2}{\Delta V_B}\frac{2\pi}{N_c(D-1)D}\;\langle \al\, G_{\mu \nu}^a G^{\mu \nu,a} \rangle_B\,.
\label{deltaEnp}
\end{align}
The leading ultraviolet divergence of the gluon condensate has been calculated in perturbation theory at 
four loops \cite{Schroder:2003uw}. The determination of the finite piece requires lattice simulations \cite{Hietanen:2004ew,Hietanen:2006rc}
and a computation to change from the lattice to dimensional regularization \cite{DiRenzo:2006nh}. 
Taking the result (in the Euclidean) from the last reference and renormalizing the bare expression in Eq.~(\ref{deltaEnp}) in the $\rm MS$ scheme yields
\begin{align}
\delta E_{s}^{np}(\nu) 
=&-\frac{C_A^3 C_F r^2 \al^4 }{\Delta V} \bigg[
\Big(\frac{43}{6} - \frac{157}{384} \pi^2 \Big) \Big( \ln \Big[\frac{\nu}{C_A \al}\Big] \nn\\
&- \frac12 (\ln (16 \pi)+ \gamma_E) - \frac{1}{8}\Big) + 2 B_G
\bigg]\,,
\label{NPlog}
\end{align}
where $B_G^{(SU(3))}=-0.2\pm 0.4({\rm MC}) \pm 0.4(\rm NSPT)$.
%
%
%
This result is of the same order as the ultrasoft four-loop contribution, i.e. ${\cal O}(r^2\al^4/\Delta V)$. 
For $\nu \equiv \Delta V$ however, the $\ln \Big[\frac{\nu}{C_A \al}\Big]$ term in Eq.~\eqref{NPlog} is parametrically dominant compared to the latter and we will include it in the numerical analysis of our results in the next section.
%
\newpage

\section{Comparison to lattice data}
\begin{figure}[ht]
\centerline{\includegraphics[width=8.2 cm]{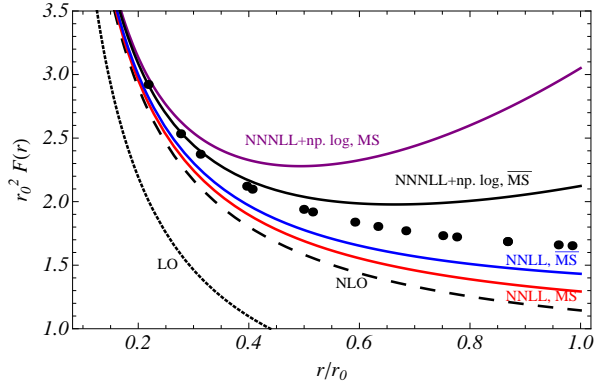}}
\caption{Plots of the analytic results for the force $F(r)=\frac{dE_s(r)}{dr}$ in ``$r_0$ units'' with $n_f=0$ in comparison to SU(2) lattice data (black dots): The dotted (LO) and dashed (LO+NLO) curves show the tree-level and one-loop results from Ref.~\cite{Schroder:1999sg}. The four other curves include in addition the new NNLO order ($\al^3 r^2$) contributions in the multipole expansion. The labels NNLL and N$^3$LL refer to the precision in the ultrasoft $\al/\Delta V$ expansion. We have plotted our results in the $\rm MS$ as well as in the $\msb$ scheme in order to make the scheme dependence visible and set $\nu=\Delta V$. Depending on the scheme we have used the values $c_{2,0}^{\rm MS}$ and $c_{2,0}^{\msb}$ given in the text for the N$^3$LL plots, which moreover incorporate the leading nonperturbative logarithm of Eq.~(\ref{NPlog}).
\label{Plot}
}
\end{figure} 
%
\nin
Now we would like to see how well the short-distance 3D lattice data can be reproduced by our theoretical 
prediction for the static singlet energy and, on the other hand, extract numerical values for $c_{2,0}$ in Eq.~\eqref{NNNLLmatching} from fits to this data.
In Fig.~\ref{Plot} we show our $\ord(\al^3 r^2)$ MS results for the static force $F(r)=\frac{dE_s(r;\nu = \Delta V)}{dr}$ up to NNLL (from Eq.~\eqref{NNLL}) and up to N$^3$LL including the leading nonperturbative logarithm (from Eq.~\eqref{NNLL} + Eq.~\eqref{NNNLLmatching} + first line of Eq.~\eqref{NPlog}) for $n_f=0$ and $N_c=2$ together with the data points from the SU(2) quenched lattice simulation of Ref.~\cite{HariDass:2007tx}.\footnote{There is also SU(3) lattice data available in Ref.~\cite{Luscher:2002qv}, but it has less points at slightly larger distances.} We use $\al=\frac{0.29}{r_0}$, where $r_0=0.5 fm$ is the Sommer scale.\footnote{For the determination of $\al$ see Refs.~\cite{OurStatPot3D,HariDass:2007tx}.} To estimate the theoretical uncertainties we also transformed the MS results to the $\msb$ scheme\footnote{Note, that this also implies $\Delta V_{\rm MS} \to \Delta V_{\msb}$ in the logs of Eq.~\eqref{NNLL} etc. giving rise to a residual scheme dependence from higher orders.} and added the corresponding curves as well as the previously known curves at LO and NLO precision to the plot. From a fit of the NNLL $\msb$ curve to the data point at the shortest distance, where we expect the best convergence of the perturbative series, we determined $c_{2,0}^{\msb}=-0.04$. Transforming this to the MS scheme gives $c_{2,0}^{MS}=2.64$.\\
If we compare the LO, NLO and the NNLO curves with ultrasoft NNLL precision, 
we find a convergent pattern, in particular in the short distance limit. 
Unlike the multipole expansion the $\al/\Delta V$ expansion does not converge well, 
even for the shortest distances that were probed on the lattice. Indeed, already at
$r/r_0 \simeq 0.22$ we have $\frac{C_A \al}{\Delta V_{\msb}} \simeq 0.60\, (\lesssim \frac{C_A \al}{\Delta V_{\rm MS}})$ for the ultrasoft expansion parameter, which is typically accompanied by the color factor $C_A$. 
The lack of convergence is in particular reflected in the huge scheme dependence of the N$^3$LL results, i.e. the big separation of the respective MS and $\msb$ curves in Fig.~\ref{Plot} at larger distances.
Therefore we do not trust in the values for $c_{2,0}$ given above and regard them instead only as a rough order of magnitude estimate. To improve on these numbers we would need lattice data at much smaller distances. 
For a more detailed numerical analysis of our results see Ref.~\cite{OurStatPot3D}.\\[-3 ex]

\section*{Acknowledgements}
\nin
This work was partially supported by the EU network contract
MRTN-CT-2006-035482 (FLAVIAnet), by the Spanish 
grant FPA2007-60275 and by the Catalan grant SGR2009-00894.





\begin{thebibliography}{}

\bibitem{FSP} W. Fischler, Nucl. Phys. {\bf B129}, 157 (1977);
Y. Schr\"oder, Phys. Lett. {\bf B447}, 321 (1999); B.A. Kniehl,
A.A. Penin, V.A. Smirnov and M. Steinhauser, Phys. Rev. {\bf D65},
091503 (2002).

\bibitem{Anzai:2009tm}
  C.~Anzai, Y.~Kiyo and Y.~Sumino,
  arXiv:0911.4335 [hep-ph].


\bibitem{Smirnov:2009fh}
  A.~V.~Smirnov, V.~A.~Smirnov and M.~Steinhauser,
  arXiv:0911.4742 [hep-ph].


\bibitem{Pineda:1997bj}
  A.~Pineda and J.~Soto,
  Nucl.\ Phys.\ Proc.\ Suppl.\  {\bf 64}, 428 (1998).

\bibitem{Brambilla:2004jw}
  N.~Brambilla, A.~Pineda, J.~Soto and A.~Vairo,
  Rev.\ Mod.\ Phys.\  {\bf 77}, 1423 (2005).


\bibitem{Appelquist:1977es}
  T.~Appelquist, M.~Dine and I.~J.~Muzinich,
  Phys.\ Rev.\  D {\bf 17}, 2074 (1978).

\bibitem{short} 
  N.~Brambilla, A.~Pineda, J.~Soto and A.~Vairo,
  Phys.\ Rev.\  D {\bf 60}, 091502 (1999).

\bibitem{Brambilla:1999xf}
  N.~Brambilla, A.~Pineda, J.~Soto and A.~Vairo,
  Nucl.\ Phys.\  B {\bf 566}, 275 (2000).

\bibitem{OurStatPot3D}
  A.~Pineda and M.~Stahlhofen,
  Phys.\ Rev.\  D {\bf 81}, 074026 (2010)


\bibitem{Schroder:1999sg}
  Y.~Schroder,
  ``The static potential in QCD'',
  DESY-THESIS-1999-021.

\bibitem{Pineda:1997ie}
  A.~Pineda and J.~Soto,
  Phys.\ Lett.\  B {\bf 420}, 391 (1998).

\bibitem{KP1} 
  B.~A.~Kniehl and A.~A.~Penin,
  Nucl.\ Phys.\  B {\bf 563}, 200 (1999).

\bibitem{Eidemuller:1997bb}
  M.~Eidemuller and M.~Jamin,
  Phys.\ Lett.\  B {\bf 416}, 415 (1998).
 

\bibitem{Brambilla:2006wp}
  N.~Brambilla, X.~Garcia i Tormo, J.~Soto and A.~Vairo,
  Phys.\ Lett.\  B {\bf 647}, 185 (2007).


\bibitem{RG} 
  A.~Pineda and J.~Soto,
  Phys.\ Lett.\  B {\bf 495}, 323 (2000).



  
  


 
   
  



  
\bibitem{Schroder:2003uw}
  Y.~Schroder,
  Nucl.\ Phys.\ Proc.\ Suppl.\  {\bf 129}, 572 (2004).
 
\bibitem{Hietanen:2004ew}
  A.~Hietanen, K.~Kajantie, M.~Laine, K.~Rummukainen and Y.~Schroder,
  JHEP {\bf 0501}, 013 (2005).
  
\bibitem{Hietanen:2006rc}
  A.~Hietanen and A.~Kurkela,
  JHEP {\bf 0611}, 060 (2006).

\bibitem{DiRenzo:2006nh}
  F.~Di Renzo, M.~Laine, V.~Miccio, Y.~Schroder and C.~Torrero,
  JHEP {\bf 0607}, 026 (2006).
  


 

\bibitem{HariDass:2007tx}
  N.~D.~Hari Dass and P.~Majumdar,
  Phys.\ Lett.\  B {\bf 658}, 273 (2008).

 
\bibitem{Luscher:2002qv}
  M.~Luscher and P.~Weisz,
  JHEP {\bf 0207}, 049 (2002).

 

\end{thebibliography}







\end{document}